\documentclass[twocolumn,english]{article}
\usepackage[top=20truemm,bottom=20truemm,left=20truemm,right=20truemm]{geometry}
\usepackage[T1]{fontenc}
\usepackage{here}
\usepackage{graphicx}
\usepackage{babel}
\usepackage{url}
\usepackage{titlesec}
\titleformat*{\section}{\large\bfseries}
\titleformat*{\subsection}{\normalsize\itshape}
\usepackage[singlelinecheck=false]{caption}
\makeatletter

\renewcommand{\maketitle}
{
\bgroup\setlength{\parindent}{0pt}
\begin{flushleft}
\textbf{\huge \@title\\}
{\normalsize \@author}
\end{flushleft}\egroup
}

\makeatother

\title{Optimizing scoring function of dynamic programming of pairwise profile alignment using derivative free neural network}
\date{}
\author{%
\vspace{1em}
Kazunori D Yamada$^{1,2\ast}$\\
\vspace{1em}
{\small
$^{1}$Graduate School of Information Sciences, Tohoku University, Sendai, Japan,
$^{2}$Artificial Intelligence Research Center, National Institute of Advanced Industrial Science and Technology (AIST), Tokyo, Japan,
\vspace{2em}
}
}

\begin{document}

\twocolumn[
\maketitle
]

\begin{abstract}
A profile comparison method with position-specific scoring matrix (PSSM) is one of the most accurate alignment methods. Currently, cosine similarity and correlation coefficient are used as scoring functions of dynamic programming to calculate similarity between PSSMs. However, it is unclear that these functions are optimal for profile alignment methods. At least, by definition, these functions cannot capture non-linear relationships between profiles. Therefore, in this study, we attempted to discover a novel scoring function, which was more suitable for the profile comparison method than the existing ones. Firstly we implemented a new derivative free neural network by combining the conventional neural network with evolutionary strategy optimization method. Next, using the framework, the scoring function was optimized for aligning remote sequence pairs. Nepal, the pairwise profile aligner with the novel scoring function significantly improved both alignment sensitivity and precision, compared to aligners with the existing functions. Nepal improved alignment quality because of adaptation to remote sequence alignment and increasing the expressive power of similarity score. The novel scoring function can be realized using a simple matrix operation and easily incorporated into other aligners. With our scoring function, the performance of homology detection and/or multiple sequence alignment for remote homologous sequences would be further improved.
\end{abstract}

\section*{Introduction}
The profile comparison alignment method with a position-specific scoring matrix (PSSM) \cite{Altschul1997} is one of the most accurate alignment methods. The PSSM is a two dimensional vector (matrix) for sequence length. Each element in the vector consists of a 20 dimensional numerical vector, in which each value represents the likelihood of the existence of each amino acid position in a biological sequence. Here, we designed the vector inside PSSM as a position-specific scoring vector (PSSV). In a profile alignment, cosine similarity or correlation coefficient is generally calculated against the PSSVs to calculate similarity or dissimilarity between the two sites in the sequences of interest on dynamic programming (DP) \cite{Tomii2004,Wu2008}. Profile alignment methods using these functions have been successful for a long time \cite{Tomii2005}, although cosine similarity or correlation coefficient cannot capture the non-linear relationship between two vectors and the similarity between two sites is not always expressed by linear relationships.\par
The performance of profile sequence alignment has been improved by various studies in the past decades. For example, HHalign improved alignment quality using profiles constructed with the hidden Markov model, which provided more information than PSSM \cite{Soeding2005}, MUSTER incorporated protein structural information in a profile \cite{Wu2008}, and MRFalign utilized the Markov random fields to improve alignment quality \cite{Ma2014}. Although various methods have been devised from different perspectives, studies to develop the scoring function itself with sophisticated technologies are lacking.\par
Neural networks are computing system, which mimic biological nervous system of animal brains. Theoretically, it can approximate any function regardless of linearity of the functions \cite{Gybenko1989}. Neural networks are attracting attention from various areas of research, including bioinformatics, due to the availability of improved computational methods and the explosive increase in available data. In recent years, these algorithms have been vigorously applied to bioinformatics. For example, several studies applied a deep neural network model to predict protein-protein interaction \cite{Du2017,Sun2017}, protein structure \cite{Wang2016,Spencer2015} and various other  biological conditions such as residue contact map, backbone angles, and solvent accessibility \cite{DiLena2012,Heffernan2017}. These algorithms basically used the backpropagation method, which requires derivation of a cost function for searching optimal parameters, and few studies implemented derivative free neural network.\par
In this study, we utilized the neural network to optimize a scoring function. In the process, we first combined two PSSVs (for which we wanted to calculate similarity) derived from two sites and set it as an input vector. A target vector was required to implement supervised learning. However, in this case, we did not have the target vector because the ideal function and an ideal similarity score for each site were unknown, and thus, the scoring function could not be directly optimized. Instead, we calculated the entire DP table for the input sequences and the difference between the resultant alignment and the correct alignment was used for calculating cost. In this case, we could not use the backpropagation method for optimal weight search because we lacked the derivation of the cost function required for this search. Namely, we could not incorporate our idea in the conventional neural network framework. Therefore, we newly utilized the covariance matrix adaptation evolution strategy (CMA-ES) \cite{Hansen1996}, which is an adaptive optimization method modifying the basic evolutionary strategy \cite{Beyer1993}, as the search method for neural network  to realize derivative free neural network calculation. Using this framework, we attempted to produce higher performance scoring function for remote sequence alignment in this study.

\section*{METHODS}
\subsection*{Dataset}
We downloaded the non-redundant subset of SCOP40 (1.75 release) \cite{Andreeva2008}, in which sequence identity between any sequence pair is less than 40\%, from ASTRAL \cite{Chandonia2004}. We selected the remote sequence subset since we wanted to improve the remote sequence alignment quality. The SCOP is a protein domain dataset where sequences are classified in hierarchical manner by class, fold, superfamily, and family. All notations of the superfamily in the dataset were sorted by alphabetical order and all superfamilies, the ordered numbers of which were multiples of three, were classified into a learning dataset, whereas the others were classified into a test dataset. We obtained 3,726 and 6,843 sequences in the learning and test datasets, respectively. Next, we randomly extracted a maximum of 10 pairs of sequences from each superfamily to negate a bias induced by different volumes of each superfamily and used these sequence pairs for subsequence construction of PSSM. We confirmed that sequences in each pair were from the same family to obtain decent reference alignment. Finally, we obtained 1,721 and 3,195 sequence pairs in the learning and test datasets.

\subsection*{Construction of profiles and reference alignments}
\begin{figure}[H]
\begin{center}
\includegraphics{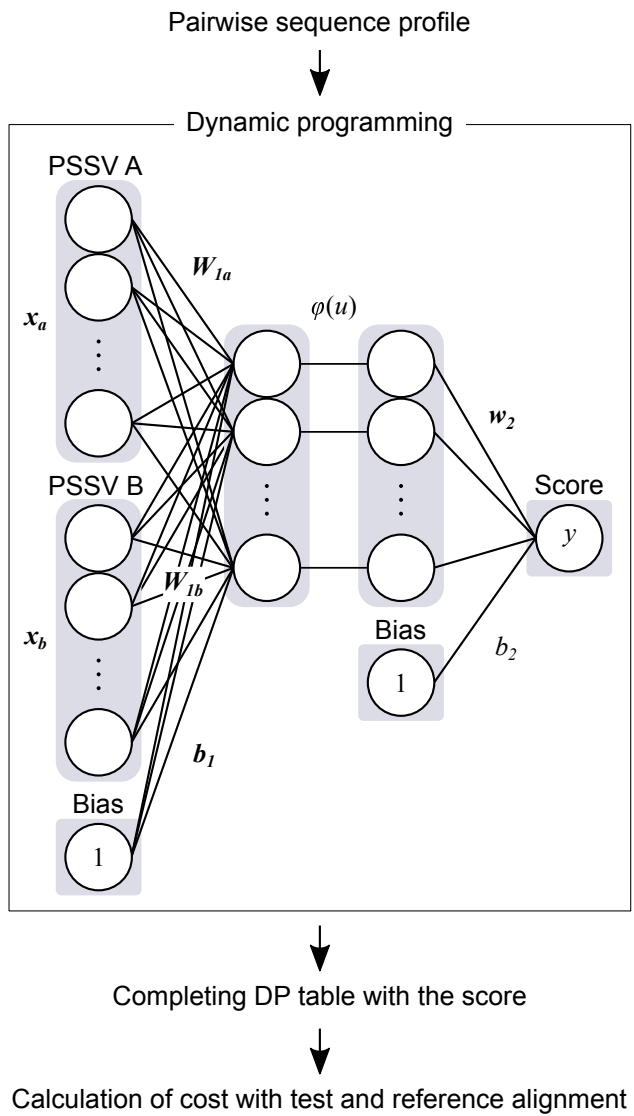}
\end{center}
\caption{Schematic diagram of learning network developed in this study. The upper case letters in italics and bold face, the lower case letters in italics and bold face, and the lower case letters in italics represent matrix, vector, and scalar values, respectively. The activating function is represented by $\phi$().}
\label{fig01}
\end{figure}

Figure \ref{fig01} shows the learning network computed in this study. We calculated similarity scores between two PSSVs using the neural network. At first, the summation of matrix products between xa (the PSSV A) and W1a, xb (the other PSSV B) and W1b, and 1 (bias) and b1 in the neural network were calculated. The resultant vector was transformed by an activating function, $\phi$(). Finally, the summation of the dot products between the transformed vector and w2, and 1 and b2 was calculated. The resultant value was used as the similarity score for the two sites. Namely, the forward calculation was computed by the following equation. Here, y is the similarity score.\par

\begin{eqnarray*}
y={\bf w_2}\phi({\bf x_a}{\bf W_{1a}}+{\bf x_b}{\bf W_{1b}}+{\bf b_1})+b_2
\end{eqnarray*}

The complete DP table was calculated using the similarity score and a final pairwise alignment was produced. The pairwise alignment and its corresponding reference alignment were compared to each other and an alignment sensitivity score, described below, was calculated. The subtraction of the alignment sensitivity score from 1 was used as cost for searching optimum weight by the neural network with CMA-ES.

We set the weights W1a and W1b equal to each other (shared weight) so that the network outputs same value even though the input order of the two PSSVs were opposite. The number of units of the middle layer was set to 144. The rectified linear unit was utilized as the activation function. We set $\sigma$, $\lambda$, and $\mu$ as 0.032, 70, and 35, respectively, as parameters for CMA-ES. Here, $\sigma$ is almost equivalent to step size of the gradient descent method, and $\lambda$ and $\mu$ indicate the number of descendant and survival individuals in evolutional process. In actual learning, we read training datasets in batch manner. The learning loop was stopped using the early stopping criteria by checking the dissociation between the training and validating curves. The initial weight was derived from parameters that mimicked the correlation coefficient. To generate the initial weight, we randomly generated 200,000 PSSM pairs and learned them using multilayer perceptron with hyperparameters (the dimension of weight and activating function) identical to the above hyperparameters. In addition to the weights, we simultaneously optimized the open and extension gap penalties. The initial values of open and extension gap penalties were set as -1.5 and -0.1.

\subsection*{Alignment algorithm}
In this study, we implemented the semi-global alignment method, namely global alignment with free end-gaps method \cite{Gotoh1982,Needleman1970}.

\subsection*{Metrics of alignment quality}
The alignment quality was evaluated using alignment sensitivity and precision \cite{Biegert2009}. The alignment sensitivity was calculated by dividing the number of correctly aligned sites by the number of non-gapped sites in a reference alignment. In contrast, alignment precision was calculated by dividing the number of correctly aligned sites by the number of non-gapped sites in a test alignment.

\subsection*{Calculation of residue interior propensity}
The relative accessible surface area (rASA) for residues of all proteins in the learning and test dataset was calculated by areaimol in CCP4 package version 6.5.0 \cite{Winn2011}. The residues of which rASA is less than 0.25 were counted as an interior residue and the other residues were counted as surface residue, according to a previous study \cite{Levy2010}. We divided the ratio of the interior residues by the background probability of residues to calculate the residue interior propensity. The residue interior propensity is the likelihood of a residue existing inside a protein. Namely, propensity greater than 1 signifies that the probability of the residue to be inside the protein is high.

\section*{RESULTS AND DISCUSSION}
\subsection*{Gap optimization of existing functions}
At first, we conducted gap penalty optimization of the existing scoring functions such as cosine similarity and correlation coefficient on the learning dataset. We computed both alignment sensitivity and precision for aligners using these functions, changing open and extension gap penalties by 0.1 increments from -2.0 to -0.6 and from -0.4 to -0.1, respectively. The best alignment sensitivity was selected as the optimum combination among the combinations of open and extension gap penalties. As shown in Table \ref{table:01}, the best gap penalty combination for cosine similarity and correlation coefficient was (-1.0, -0.1) and (-1.5, -0.1).

\subsection*{Optimization of scoring function of the neural network}
Next, we conducted optimization of scoring function on the neural network with CMA-ES. During learning, we randomly divided the learning dataset into two subsets, namely, the training and validation datasets, which included 1,536 and 160 pairwise PSSV sets and its corresponding reference alignments as targets, respectively. Since calculation of CMA-ES in our parameter settings requires more than 100,000 times DP (the size of training dataset $\times$ $\lambda$) per epoch, the consumption of computer resources was large and calculation time was long even when 24 threads were used with the C++ program; therefore, we set the maximum limit for epoch to a small number such as 150. We selected the best scores from the validation scores of the last fifth part of an entire epoch (which was derived from 145th epoch) and obtained final weight and bias matrices, namely, the substance of a novel scoring function and optimal gap penalty combination, respectively. As a result, optimal combination of open and extension gap penalty for the final weight and bias matrix were approximately -1.7 and -0.2.\par
Finally, we implemented the pairwise profile aligner with the weight and bias matrices as novel scoring function and named it as neural network enhanced profile alignment library (Nepal). Our aligner and scoring function (weight and bias matrices) can be downloaded from https://github.com/yamada-kd/nepal.

\begin{table}[b]
{\footnotesize{
\caption{Gap optimization of the existing scoring function}
\label{table:01}
\begin{tabular*}{\columnwidth}{@{\extracolsep{\fill}}lcccc}\hline
       & Open & Extension & Sensitivity & Precision\\ \hline
Cosine & -1.0 & -0.1      & 0.6837      & 0.6550\\
CC     & -1.5 & -0.1      & 0.6882      & 0.6613\\ \hline \vspace{1pt}
\end{tabular*}
{Open and Extension indicate optimized open and extension gap penalties, respectively, and Cosine and CC represent aligners with a cosine similarity and correlation coefficient as scoring functions, respectively.}
}}
\end{table}

\subsection*{Benchmark of Nepal and other aligners with existing function on the test dataset}
Next, we conducted benchmark test of Nepal and other aligners with existing functions on the test dataset. In addition to profile comparison methods, we examined the performance of sequence comparison aligners with difference substitution matrices such as BLOSUM62 \cite{Henikoff1992} and MIQS \cite{Yamada2014} for reference. We used -10 and -2 as open and extension gap penalties, respectively, based on a previous study \cite{Yamada2014}. When calculating alignment qualities, the test dataset was further categorized into remote and medium subset depending on pairwise sequence identity of the reference alignments. The remote and medium subset includes sequence pairs, of which each sequence identity was not lower than 0\% and less than 20\%, and not lower than 20\% and less than 40\%, respectively. Generally, a pairwise alignment between sequences of lower identity such as those in the twilight zone is more difficult \cite{Rost1999}.\par
Table \ref{table:02} shows alignment quality scores for each method. Results show that among the existing methods, including sequence comparison methods, the method with the best performance from all perspectives was the profile comparison method with correlation coefficient scoring function. In contrast, Nepal improved both alignment sensitivity and precision compared to this method. Actually, these improvements were statistically significant according to Wilcoxon signed rank test with Bonferroni correction even when significance level ($\alpha$) is set to 0.01. Comparison between sequence-based methods with different substitution matrices such as MIQS and BLOSUM62 showed that the gain of improvement of MIQS compared to BLOSUM62 was more significant for the remote subset than the medium subset. This was expected since MIQS was originally developed to improve remote homology alignment. This trend was observed regarding the relationship between Nepal and correlation coefficient implemented aligner, where Nepal improved both alignment sensitivity and precision by about 4\% and 1\% in remote and medium subsets, respectively. This indicated that the novel scoring function was optimized for remote sequence alignment. This is expected because sequence alignment between sequences with closer identities was easier than those with remote identities. Therefore, during optimization, the novel scoring function would be optimized to be naturally advantageous for remote sequence alignments. Since the problem regarding remote relationship holds true for sequence similarity search \cite{Yamada2014,Fariselli2007}, the novel scoring function of our method could be useful for improving the performance of remote similarity search methods.

\begin{table}[h]
{\footnotesize{
\caption{Comparison of Nepal with other alignment methods}
\label{table:02}
\begin{tabular*}{\columnwidth}{@{\extracolsep{\fill}}lccc}\hline
           & Remote        & Medium        & All\\
           & [0,20)        & [20,40)       & [0,40)\\
           & (1,405 files) & (1,790 files) & (3,195 files)\\ \hline
\it Sensitivity\\
\,\,Nepal  & \bf 0.5317              & \bf 0.8343              & \bf 0.7012\\
\,\,Cosine & ~~~~0.5045$^{\ast\ast}$ & ~~~~0.8246$^{\ast\ast}$ & ~~~~0.6838$^{\ast\ast}$\\
\,\,CC     & ~~~~0.5135$^{\ast\ast}$ & ~~~~0.8269$^{\ast\ast}$ & ~~~~0.6891$^{\ast\ast}$\\
\,\,MIQS   & ~~~~0.2775$^{\ast\ast}$ & ~~~~0.7316$^{\ast\ast}$ & ~~~~0.5319$^{\ast\ast}$\\
\,\,BL62   & ~~~~0.2333$^{\ast\ast}$ & ~~~~0.6955$^{\ast\ast}$ & ~~~~0.4923$^{\ast\ast}$\\
\it Precision\\
\,\,Nepal  & \bf 0.5031              & \bf 0.8102              & \bf 0.6751\\
\,\,Cosine & ~~~~0.4753$^{\ast\ast}$ & ~~~~0.7999$^{\ast\ast}$ & ~~~~0.6571$^{\ast\ast}$\\
\,\,CC     & ~~~~0.4858$^{\ast\ast}$ & ~~~~0.8032$^{\ast\ast}$ & ~~~~0.6636$^{\ast\ast}$\\
\,\,MIQS   & ~~~~0.2654$^{\ast\ast}$ & ~~~~0.7134$^{\ast\ast}$ & ~~~~0.5164$^{\ast\ast}$\\
\,\,BL62   & ~~~~0.2317$^{\ast\ast}$ & ~~~~0.6902$^{\ast\ast}$ & ~~~~0.4885$^{\ast\ast}$\\  \hline \vspace{1pt}
\end{tabular*}
{The interval on the second line in the header represents sequence identity (\%) of each division. Methods such as Cosine, CC, MIQS, and BL62 indicate profile comparison methods with cosine similarity and correlation coefficient and  sequence comparison methods with MIQS and BLOSUM62. The double asterisks on the score ($^{\ast\ast}$) indicate p-value $<$ 0.01 on Wilcoxon signed rank test with Bonferroni correction when the method is compared to Nepal.}
}}
\end{table}

\subsection*{Importance of attributes using the connection weight method}
\begin{figure}[h]
\begin{center}
\includegraphics{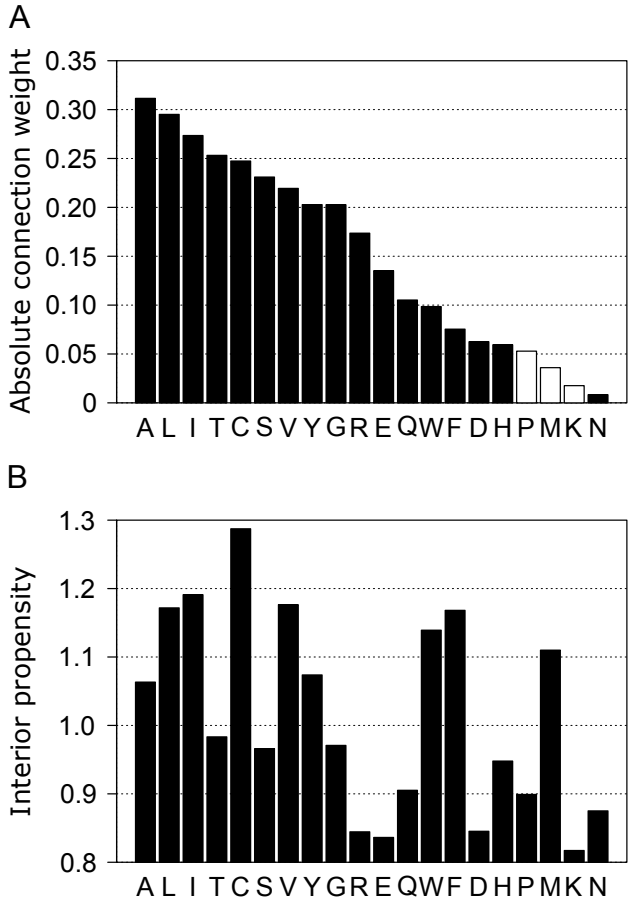}
\end{center}
\caption{(A) Absolute connection weight for each attribute, which corresponds to the profile value of each amino acid. Filled and open bar represents positive and negative sign of original connection weight, respectively. (B) The residue interior propensity against whole data in the study.}
\label{fig02}
\end{figure}

Finally, we calculated the importance of 20 attributes using the connection weight method \cite{Olden2004}. As shown in Figure \ref{fig02}A, the connection weights against each attribute, namely each amino acid, were distributed to various values. This indicated that our developed scoring function discerned the importance of the attributes depending on the variety of amino acids.

According to the results, the connection weight of hydrophobic residues such as Leu, Ile, and Val were of higher value. These residues are located mostly inside the hydrophobic cores of proteins. In addition, as shown in Figure \ref{fig02}B, the other residues which also tend to locate inside proteins, such as Ala, Cys, and Tyr, were of higher importance. In contrast, residues which tend to locate on protein surface, such as Asp, Pro, Lys, and Asn, were of lower importance. The Spearman's rank correlation coefficient between the connection weight and interior propensity was approximately 0.6 and the value was statistically significant (p-value $<$ 0.05). While residues which are exposed on the protein surface are subject to higher mutation pressures, interior residues are less susceptible to mutation \cite{Zhang2015}. This is because the protein structure is disrupted if mutations in the interior residues collapse the hydrophobic core  \cite{Chakravarty1999}. The scoring function constructed in this study was optimized for alignment of remote homologous sequences. According to the previous study based on substitution matrices \cite{Kinjo2004}, hydrophobicity of residues  was the dominant property of remote sequence substitution rather than simple mutability. This fact partially represents that for remote sequence alignment, residues occupying interior locations in a protein higher order structure with less susceptibility to mutation pressure are considered more meaningful. Since our scoring function was also optimized for remote sequence alignment, the above property would be observed and this fact paradoxically suggests that our scoring function was optimized for remote sequence alignment. Collectively, this property is one of the reasons for the superiority of our method to the existing ones.\par
\begin{figure}[h]
\begin{center}
\includegraphics{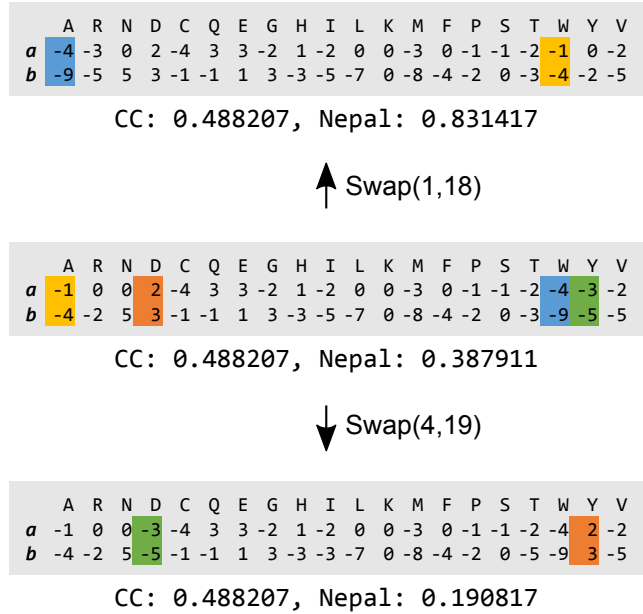}
\end{center}
\caption{Transition of similarity scores depending on site swapping. In each panel, a and b stand for PSSV A and B respectively. The middle panel represents an original PSSV and similarity scores calculated by correlation coefficient (CC) and Nepal. The top and bottom panel stands for the resultant PSSVs and the similarity scores.}
\label{fig03}
\end{figure}

In addition, although the connection weight consisted of various values, it would at least contribute to increasing the expressive power of the novel scoring function. For example, we wanted to calculate the similarity score between PSSV A (a) and B (b) as shown in Figure \ref{fig03}. The original scores are 0.488207 and 0.387911 when calculated using the correlation coefficient and Nepal score, respectively, (middle panel Figure \ref{fig03}). The scores calculated by correlation coefficient did not change when the 1st and 18th sites or the 4th and 19th sites were swapped. This was unexpected since the converted PSSV obtained after swapping was not identical to the original one. This could be one of the drawbacks of using unweighted linear function such as cosine similarity and correlation coefficient. In contrast, Nepal scores changed after the swapping, which varied with the change in PSSV. Actually, there were about 290,000 overlaps when we calculated similarity score to six places of decimal against randomly generated one million PSSVs using correlation coefficient, whereas there were approximately 180,000 overlaps when Nepal was used. These overlaps would negatively affect DP computation because higher overlap scores would cause difficulty in deciding the correct path, especially during the computation of maximum three values derived from up, diagonal, and left side of the DP cell.\par
Collectively, the different weights based on amino acid variety presented by the connection weight method is one of the reasons why Nepal score improved the alignment quality compared to the existing scoring functions. 

\subsection*{CONCLUSION}
In this study, we developed a new derivative free neural network with CMA-ES. Using this framework, we developed a novel scoring function for profile comparison and Nepal, a pairwise profile aligner with the scoring function. Large computational resources were required by our learning procedure with the derivative free neural network; thus, we could not examine whether the learning was converged enough because of our limited computational environment. Nevertheless, Nepal significantly improved alignment quality of profile alignment, especially for alignment of remote relationships, compared to the existing scoring functions. Nepal improved alignment quality because of adaptation to remote sequence alignment and increasing the expressive power of similarity score. The novel scoring function can be realized using a simple matrix operation and the parameters are provided on https://github.com/yamada-kd/nepal. In future, the performance of distant homology detection method or that of multiple sequence alignment method for remote homologous sequences may be further improved with our scoring function.

\section*{ADDITIONAL INFORMATION}
\subsection*{Acknowledgements}
We are grateful to Dr Kentaro Tomii, Dr. Satoshi Omori and Mr. Tsukasa Nakamura for constructive discussion. Computations were partially performed on the NIG supercomputer at the ROIS National Institute of Genetics.

\subsection*{Funding}
This work was supported in part by the Top Global University Project from the Ministry of Education, Culture, Sports, Science, and Technology of Japan (MEXT)

\subsection*{Availability of data and material}
The source code of Nepal and the learned parameters are available at GitHub (https://github.com/yamada-kd/nepal).

\subsection*{Abbreviations}
CMA-ES: covariance matrix adaptation evolution strategy; DP: dynamic programming; PSSM: position-specific scoring matrix; PSSV: position-specific scoring vector

\subsection*{Competing interests}
The authors declare that they have no competing interests.

\subsection*{Author Contribution}
KDY did everything.

{\footnotesize
\bibliographystyle{unsrt}
\bibliography{nepal}
}
\end{document}